\documentclass[aps,prl,reprint,showpacs,preprintnumbers, amsmath,amssymb,floatfix]{revtex4-2}

\usepackage{graphicx}
\usepackage{dcolumn}
\usepackage{bm}
\usepackage{hyperref}
\usepackage[normalem]{ulem}

\usepackage{color}
\usepackage{subfigure}
\definecolor{niceblue}{rgb}{0.388235, 0.627451, 0.847059}
\definecolor{nicered}{rgb}{0.7,0.1,0.1}
\definecolor{nicegreen}{rgb}{0.1,0.5,0.1}
\definecolor{JMBcomment}{RGB}{64, 214, 136}

\begin{document}

\preprint{N3AS-21-008}

\title{Neutron star structure with a new force between quarks}

\author{Jeffrey M. Berryman}
\email{jeffberryman@berkeley.edu}
\affiliation{Department of Physics, University of California, Berkeley, CA~94720, USA}
\affiliation{Department of Physics and Astronomy, University of Kentucky, Lexington, 
KY~40506-0055, USA}

\author{Susan~Gardner}
\email{gardner@pa.uky.edu}
\affiliation{ Department of Physics and Astronomy, University of Kentucky, Lexington, KY 40506-0055, USA}


\begin{abstract}
The discovery of nondiffuse sources of gravitational waves through compact-object mergers opens new prospects for the study of physics beyond the Standard Model. In this paper, we study the effects of a new force between quarks, suggested by the gauging of baryon number, on pure neutron matter at supranuclear densities. This leads to a stiffening of the equation of state, allowing neutron stars to be both larger and heavier and possibly accommodating the light progenitor of GW190814 as a neutron star. The role of conventional three-body forces in neutron star structure is still poorly understood, though they can act in a similar way, implying that the mass and radius do not in themselves resolve whether new physics is coming into play. However, a crucial feature of the scenario we propose is that the regions of the new physics parameter space that induce observable changes to neutron star structure are testable at low-energy accelerator facilities. This testability distinguishes our scenario from other classes of new phenomena in dense matter.
\end{abstract}

\maketitle

{\it Introduction.} 
The environment of a proto-neutron star (NS) has long been known to be exquisitely sensitive to the appearance of light, new physics (NP), such as axions~\cite{Iwamoto1984PhRvL..53.1198I, Turner1988PhRvL..60.1797T, Brinkmann1988PhRvD..38.2338B, Burrows1989PhRvD..39.1020B, Janka1996PhRvL..76.2621J, Hanhart:2000ae, Sedrakian2016PhRvD..93f5044S} or dark photons~\cite{Dent2012arXiv1201.2683D,Dreiner2014PhRvD..89j5015D,Kazanas2015NuPhB.890...17K}, through cooling effects. Dark matter can also be captured by NSs, altering them so severely that the established existence of NSs constrains dark matter properties~\cite{Gould:1987ir,McDermott:2011jp}. With the advent of gravitational wave (GW) observations of compact object mergers~\cite{LIGO2016PhRvL}, new windows on the nature of matter at supranuclear densities open~\cite{Llanes-Estrada2019PrPNP.10903715L}. There has been much discussion of emergent phenomena within QCD at near-zero temperature $T$ with neutron chemical potential $\mu_n$, with transitions to condensed phases~\cite{Glendenning:1992vb} of pions~\cite{Bahcall1965PhRv..140.1445B} and kaons~\cite{Kaplan1986PhLB..175...57K}, or to a spin-color-flavor-locked $qq$ phase~\cite{Rajagopal2001afpp.book.2061R, Alford2001ARNPS..51..131A}, or to states with substantial admixtures of $s$ and $\bar s$ quarks all possible~\cite{Glendenning:1984jr, Alcock1986ApJ...310..261A, weber2005PrPNP..54..193W}. The structure of NSs is also sensitive to new neutron decay channels, as noted in new physics models for the neutron lifetime anomaly~\cite{Fornal2018PhRvL.120s1801F, Wietfeldt2011RvMP...83.1173W}, yielding severe constraints~\cite{Mckeen2018PhRvL.121f1802M, Motta2018JPhG...45eLT01M, Baym2018PhRvL.121f1801B}. 

Here, we consider minimal extensions of the Standard Model (SM) that give rise to new, short-range interactions between quarks. In particular, we consider U(1)$_X$ extensions that couple to baryon number $B$; such models have proven popular in searches for light hidden sectors in low-energy accelerator experiments~\cite{Alexander2016arXiv160808632A} because the possibility that the neutrino has a Majorana mass predicates that the $B-L$ symmetry of the SM is broken. If $B$ symmetry is spontaneously broken to give a gauge boson $X$ no lighter than a few hundred MeV, then the new interaction is largely shielded from constraints from low-energy experiments. In particular, its contribution to the nucleon-nucleon ($NN$) force can be hidden within the short-distance repulsion of the phenomenological $NN$ force in the SM, recalling, e.g., the repulsive hard core of the Reid potential at separations of $r_{\rm hc}=0.5\,{\rm fm}$~\cite{Reid1968AnPhy..50..411R}, yet it can modify the the neutron matter equation of state (EoS) at supranuclear densities, i.e., beyond the saturation number density of ordinary nuclear matter, $n_{\rm sat}$~\footnote{In effective field theory (EFT) language, its effect is embedded within a low-energy constant at renormalization scales below $\mu < M_X$, but $X$ becomes an active degree of freedom at scales above $\mu > M_X$. In the current context we suppose $X$ is explicit if $k_F c$ is not grossly smaller than $M_X$}. We expect these models to be accompanied by electromagnetic signatures, such as, e.g., brighter kilonovae, due to $X-\gamma$ mixing, but reserve this for later work~\cite{UsUpcoming}.

{\it Theoretical Framework.}
The lighter compact object in GW190814 is of $2.50 - 2.67 M_\odot$ (90\% credible level) in mass~\cite{Abbott:2020khf}, and is likely too heavy to be a NS, at least within a nonrelativistic many-body approach using $NN$ forces from chiral effective theory, with low-energy constants (LECs) determined from nuclear data~\cite{Abbott:2020khf}. Relativistic mean-field models can generate masses in excess of 2.6 $M_\odot$~\cite{Fattoyev:2020cws}, though they are challenged by constraints from heavy-ion collisions (HIC)~\cite{Danielewicz:2002pu}; we note 
Refs.~\cite{Zhang:2020zsc, Tsokaros:2020hli, Tews:2020ylw, Dexheimer:2020rlp, Huang:2020cab, Sedrakian:2020kbi, Cao:2020zxi, Zhang:2020jmb, Das:2020dcq, Demircik:2020jkc, Thapa:2020ohp, Bombaci:2020vgw, Biswas:2020xna, Li:2020ias, Roupas:2020nua, Thapa:2020usm, Rather:2020lsg, Kanakis-Pegios:2020kzp, Blaschke:2020vuy, Rather:2021yxo, Ayriyan:2021prr, Rather:2021azv} for further discussion of the light progenitor of GW190814 as a neutron star. We consider our NP model within a nonrelativistic many-body framework. Drischler \emph{et al.}~\cite{Drischler:2020fvz} recognize the importance of relativistic corrections but also think that knowledge of the high-density EoS is likely inadequate. In particular, they adopt a piecewise EoS: beyond some density cutoff, the EoS is given by the stiffest form allowed by causality. They choose a cutoff density in the region $(1-2)n_{\rm sat}$ and claim that the modified EoS and NS outcomes below that cutoff can be made without relativistic corrections. 

The chiral effective theory approach uses $NN$ and nuclear data to determine the LECs, with independent two- and three-body forces coming into play~\cite{Friar:1998zt,Epelbaum:2005bjv}. In contrast, our Abelian NP model directly yields two-body forces only. To sharpen the distinction between SM and NP effects, we employ the AV18 $NN$ interaction~\cite{Wiringa:1994wb}, whose properties are determined by $NN$ observables only.
Studies of the AV18 interaction in pure neutron matter (PNM) with different nonrelativistic methods show that it compares favorably with other interactions up to about 4$n_{\rm sat}$~\cite{Piarulli:2019pfq}. Here, we compare computations of the PNM EoS using Brueckner-Hartree-Fock (BHF) theory with the AV18 interaction with and without NP. 

{\it Secret Interactions of Quarks.}
That new interactions could exist between quarks is a long-standing possibility~\cite{Lee:1955vk,Okun:2006eb,Nelson:1989fx,Dobrescu:2014fca,GardnerHolt2016PhRvD..93k5015G}, and SM extensions in which a new vector mediator couples to baryon number $B$~\cite{Nelson:1989fx,Bailey:1994qv,Carone1995PhRvL..74.3122C,FileviezPerez2010PhRvD..82a1901F,Tulin:2014tya,Dobrescu:2014fca} yield a repulsive interaction between baryons. Thus we suppose a quark $q$ can interact via 
\begin{equation}
	{\cal L} \supset \frac{1}{3} g_X {\bar q} \gamma_\mu q X^\mu \,.
	\label{newintq}
\end{equation}
Since quarks themselves carry electric charge, this generates $X-\gamma$ mixing at the quantum level. At GeV-scale energies, we have 
\begin{equation}
	{\cal L} \supset g_B {\bar N}\gamma_\mu N X^\mu + \varepsilon e \bar N \gamma_\mu \frac{(1 + \tau_3)}{2}N X^\mu 
-\varepsilon e \bar\ell_i \gamma_\mu \ell_i X^\mu \,,
	\label{newinth}
\end{equation}
where $N$ is the nucleon doublet, $\ell_i$ is a charged lepton, and $\varepsilon$, although crudely $\approx e g_B/(4\pi)^2$, can be made smaller still~\cite{Carone1995PhRvD..52..484C}, with $\varepsilon \sim 10^{-8} - 10^{-2}$~\cite{Bjorken:2009mm}. Experimental constraints are such that the new mediator cannot be too light~\cite{Nelson:1989fx,Tulin:2014tya}; we focus on gauge mediators of about $0.2-1$ GeV in mass, for which $g_B$ can be as large as $g_B \approx 0.4$~\cite{Tulin:2014tya,Dobrescu:2014fca,Michaels:2020fzj}. However, if we adopt a U(1)$_{B_1}$ model, so that the gauge boson couples to first-generation quarks only, constraints from hadronic $J/\psi$ and $\Upsilon$ decays are weakened as $X$ can only mediate these decays through kinetic mixing and $g_B$ can be ${\cal O}(1)$. Although $\varepsilon \ne 0$, dark photon searches limit $\varepsilon \lesssim 10^{-3}$ for $M_X\sim 1\, \rm GeV$~\cite{LHCbdarkphoton2017arXiv171002867L}; we neglect all ${\cal O}(\varepsilon)$ effects in this study. We emphasize that a U(1)$_{B_{1}}$ model with the new interaction of Eq.~(\ref{newinth}) alone is incomplete: the addition of new fermions is required to make the theory consistent at high energies. Different completions are possible; we note Refs.~\cite{Foot1989PhRvD..40.2487F,He1990PhRvD..41.1636H,Carone1995PhRvD..52..484C,Dobrescu:2014fca,Feng:2016ysn} as examples. Other U(1)$_{\rm X}$ models~\cite{HeJoshiVolkas1991PhRvD..43...22H,Dobrescu:2014fca} could give rise to Eq.~(\ref{newintq}) and thus be operative. For example, if $X =(B-L)_1$, then the theory is consistent if just one right-handed neutrino is added. The possibilities can be distinguished through direct or hidden sector searches at accelerator facilities, or through $\nu$, $e^\pm$, or 
$\mu^\pm$ emission in compact object mergers.

{\it New $NN$ Forces in Neutron and Nuclear Matter.}
To evaluate the effect of the new $NN$ force in the nuclear medium and its modification to the EoS, we compute the effective two-particle interaction using the Brueckner-Bethe-Goldstone equation,
\begin{equation}
	G = V + \frac{VQ}{\omega - H_0}G,
	\label{eq:Gmatrix}
\end{equation}
with $V$ the vacuum two-nucleon potential, $Q$ the Fermi operator, $\omega$ the initial two-particle energy, and $H_0$ the in-medium Hamiltonian. The $G$-matrix is then used to calculate the single-particle potential $U(\mathbf{k})$ and the energy per nucleon $E/A$ of the system. Both $\omega$ and $H_0$ in Eq.~\eqref{eq:Gmatrix} depend on the potential $U(\mathbf{k})$; one must ensure consistency between the potential used as input and that calculated from $G$, noting Refs.~\cite{HaftelThesis,Haftel:1970zz,HjorthJensen:1995ap,UsUpcoming} for details. 

We write the two-particle potential as $V = V_\text{AV18} + V_\text{NP}$, where $V_\text{AV18}$ is that in Ref.~\cite{Wiringa:1994wb,argonne} and $V_\text{NP}$ is generated by NP, being of Yukawa form: 
\begin{equation}
    V_\text{NP} = \frac{\alpha_{B_1}}{r} e^{-M_X r},
\end{equation}
with $\alpha_{B_1} = g_{B_1}^2/4\pi$ the strength of the gauged U(1)$_{B_1}$ interaction and $M_X$ the mass of the associated boson. This interaction only ever stiffens the EoS. If the new boson is heavier than the pion and not too strongly coupled, then its effects on nuclear matter are expected to be subdominant to those of the strong interactions, comprising at most some part of the empirical LECs. For simplicity we approximate NSs as composed of PNM, noting that Table XI of Ref.~\cite{Akmal:1998cf} shows that the maximum NS masses in PNM and $\beta$-stable matter differ by less than $\approx0.5\%$. We evaluate Eq.~\eqref{eq:Gmatrix} by decomposing it into its partial wave components and summing these contributions to the EoS. We consider partial waves up to $J_\text{max} = 11$, facilitating comparison with Ref.~\cite{Piarulli:2019pfq}. Considering only the AV18 potential, we obtain $E/A = 13.7$ MeV at $n = n_\text{sat} = 0.16$ fm$^{-3}$, compared to 13.4 MeV~\cite{Piarulli:2019pfq}; for $n=0.3$ fm$^{-3}$, we obtain 26.0 MeV, in perfect agreement.

A primary source of uncertainty in nuclear matter calculations is the precise many-body technique employed. For instance, Ref.~\cite{Piarulli:2019pfq} contrasts several such methods and finds that the in-medium potential energies may differ by a factor of $\approx2$ at supranuclear densities for the same $NN$ potential. However, our purpose is to identify the allowed region of the $g_B$--$M_X$ parameter space in which the new interaction produces significant changes to NS structure. We assume that the answer to this question does not depend on the many-body method we use~\footnote{For reference, we note that the study of Ref.~\cite{Piarulli:2019pfq} shows the many-body method variation is far smaller for the simpler AV6$^\prime$ force.}, and we emphasize that the uncertainties from three-body forces in this case are moot.

\begin{figure*}[htp]
	\centering
	\subfigure[]{\includegraphics[scale=0.50]{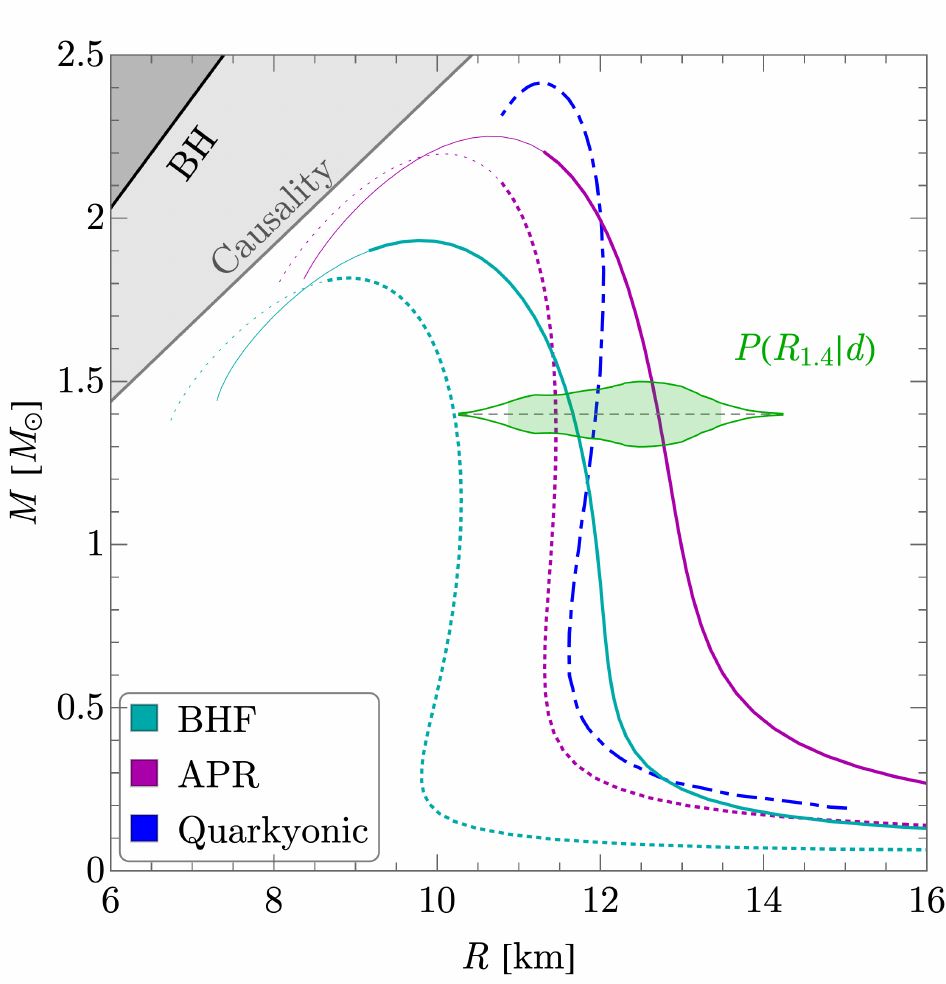}}
	\quad
	\subfigure[]{\includegraphics[scale=0.50]{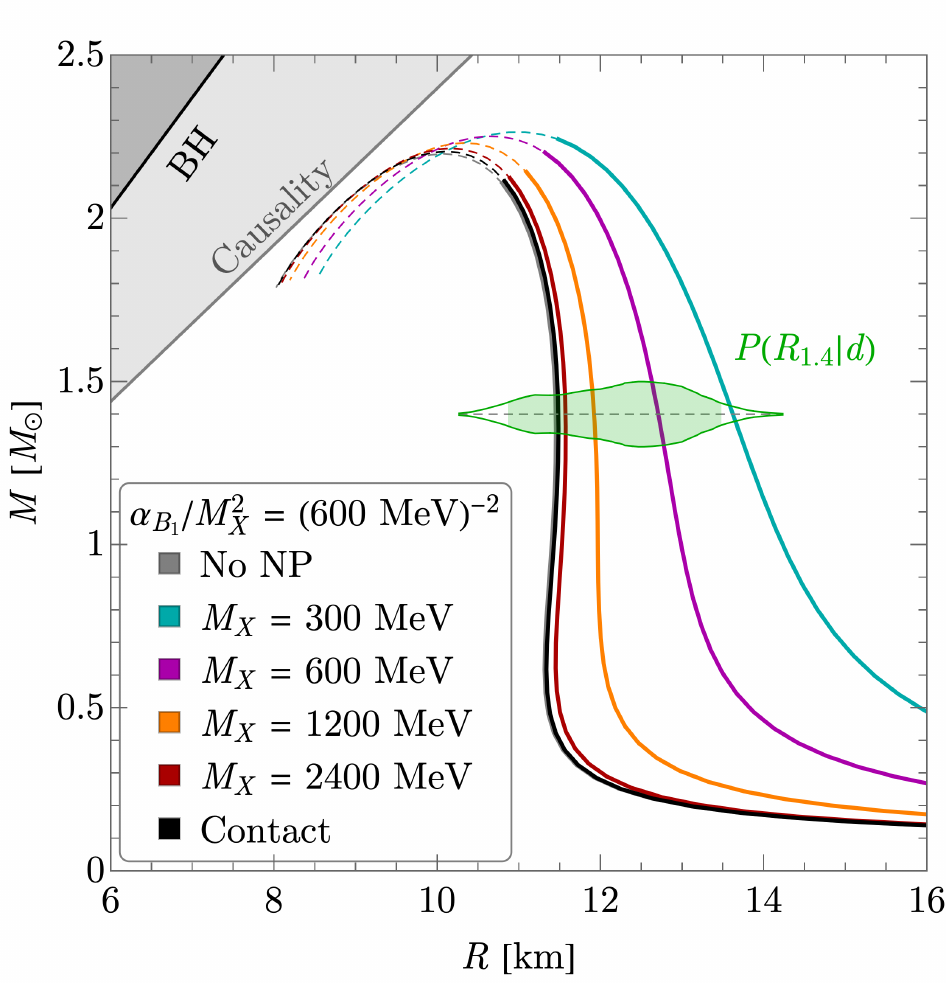}}
	\caption{(a) Mass-radius curves for our candidate EoS. Dotted lines assume only AV18 interactions \cite{Wiringa:1994wb}; solid lines introduce NP with $\alpha_{B_1} = 1$ and $M_X$ = 600 MeV. The dot-dashed line represents the quarkyonic EoS of Ref.~\cite{McLerran:2018hbz}. (b) Mass-radius curves for fixed $\alpha_{B_1}/M_X^2 = (\text{600 MeV})^{-2}$ for different values of $M_X$. In either panel, the violin region represents the PSRs+GWs+NICER posterior on $R_{1.4}$ after Fig.~10 of Ref.~\cite{Landry:2020vaw}; the shading indicates the 90\% credible region.}
	\label{fig:MR_plot}
\end{figure*}

{\it Results.} In Fig.~\ref{fig:MR_plot}a, we show the NS mass-radius ($M$-$R$) relationship of our BHF EoS; the dotted line considers only AV18, whereas the solid line introduces NP with $\alpha_{B_1} = 1$ and $M_X = 600$ MeV. Thick lines indicate where the EoS is \emph{causal}, i.e., the points for which the sound speed $c_s^2 \leq 1$ in the core; thin lines indicate where it is not. The maximum stable, causal NS mass, $M_\text{TOV}$, without NP is $\approx 1.8 M_\odot$ \footnote{The uncertainties on and correlations between the parameters of the AV18 potential have not been provided. Therefore, we are 
unable to quantify the 
uncertainties on any predicted quantities.}; even with NP, this EoS is too soft to comfortably accommodate known heavy pulsar masses \cite{Demorest:2010bx, Antoniadis:2013pzd, Arzoumanian:2017puf, Cromartie:2019kug}, $2.0-2.2 M_\odot$. This is unsurprising, since this treatment neglects relativistic corrections and three-body forces. To remedy this, we have also considered the EoS presented in Ref.~\cite{Akmal:1998cf} (``APR"), which is built on the AV18 two-nucleon potential and the Urbana IX three-body potential \footnote{We have considered the high-density phase EoS presented in this work, which includes the effects of neutral pion condensation. This phase dominates the structure of NSs with $M \gtrsim 0.5 M_\odot$; we employ it universally in our analysis.}. We approximate NP effects on this EoS by calculating the difference between $E/A$ with NP and without it in the BHF scheme, and then adding this difference to the nominal APR EoS for PNM. The results are also shown in Fig.~\ref{fig:MR_plot}a; the curves have the same interpretation as for the BHF case. Here we find $M_\text{TOV} \sim 2.1 M_\odot$ --- this is consistent with the mass of PSR J0740+6620, $2.14^{+0.10}_{-0.09} M_\odot$ \cite{Cromartie:2019kug}, but is still too light to explain GW190814. In both cases, the new interaction generates larger NS masses, as anticipated, but also increases the radius of NSs of a given mass. 

To contextualize Fig.~\ref{fig:MR_plot}a, we show the region in the $M$-$R$ plane in which the NS becomes a black hole (BH), as well as a limit on NS properties from causality~\cite{Koranda:1996jm}. Moreover, we show the posterior probability on the radius of a $1.4M_{\odot}$ NS, $P(R_{1.4}|d)$, conditioned on data $d$ from heavy pulsars, gravitational wave events and NICER observations of PSR J0030+0451, adapted from Fig.~10 of Ref.~\cite{Landry:2020vaw}. Lastly, we have included the EoS derived for quarkyonic matter in Ref.~\cite{McLerran:2018hbz} as an example with a quark/hadron QCD phase transition.The salient feature of this EoS is that the transition from hadronic to mixed hadron-quark matter induces a spike in $c_s^2$ at a few times saturation density; similar features are present in some of the sound speed profiles considered in Ref.~\cite{Tan:2020ics}. This rapid stiffening of the EoS is sufficient to allow for heavier NSs and is consistent with inferences of $c_s^2$ \cite{Landry:2020vaw} but is not strictly required. The new vector interaction induces a milder increase in $c_s^2$; the effect is that more severe increases in $M_\text{TOV}$ are correlated with larger low-mass NSs. This prediction may be crucial for disentangling the presence of NP from critical QCD phenomena.

In Fig.~\ref{fig:MR_plot}b, we fix the interaction strength to be $\alpha_{B_1}/M_X^2 = (\text{600 MeV})^{-2}$ for four finite values of $M_X$. Also shown are the contact-interaction limit and the baseline APR EoS. Lighter states generate larger contributions to the EoS, and thus to NS properties, owing largely to their effects on higher partial-wave potentials. Contact interactions only contribute to $s$-wave scattering, whereas finite-mass states contribute at all orders. Because higher partial waves become more important at higher densities, the outcome is shown in Fig.~\ref{fig:MR_plot}b. The pure contact interaction produces a result that is barely distinguishable from the nominal APR curve.

\begin{figure}[!b]
    \includegraphics[width=\columnwidth]{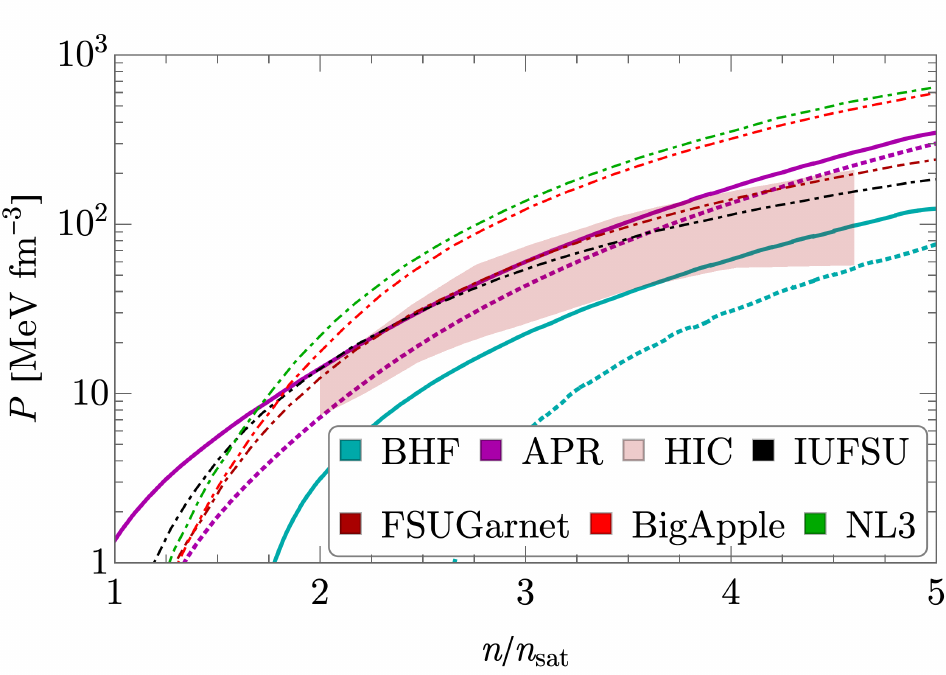}
    \caption{A comparison between the pressures in our candidate EoSs for SNM and measurements from HIC data~\cite{Danielewicz:2002pu}. Dotted lines are for AV18 interactions; solid lines introduce NP with $\alpha_{B_1} = 1$ and $M_X = 600$ MeV. We contrast these with mean-field calculations in Ref.~\cite{Fattoyev:2020cws} as dot-dashed lines.}
    \label{fig:SNM}
\end{figure}

We also calculate the EoS of symmetric nuclear matter (SNM) using the same techniques with $J_\text{max} = 8$. In Fig.~\ref{fig:SNM}, we show the pressure determined in our BHF and APR schemes, and we include NP with $\alpha_{B_1} = 1$ and $M_X = 600$ MeV. We compare these with inferences of the EoS from HIC, shown in shading. As with PNM, the pure BHF EoS is too soft to accommodate observations, but the APR EoS is a plausible candidate. The new interaction stiffens the EoS, as expected, but not so much that the HIC constraint is violated. For context, we show several mean-field EoS \cite{Lalazissis:1996rd,Fattoyev:2010mx,Chen:2014mza,Fattoyev:2020cws}. We note that the pressure for APR+NP is nonzero at empirical saturation density; we expect this to be resolved with a more refined treatment, while still generating nontrivial effects at higher densities.

\begin{figure}[!t]
    \centering
    \includegraphics[width=\columnwidth]{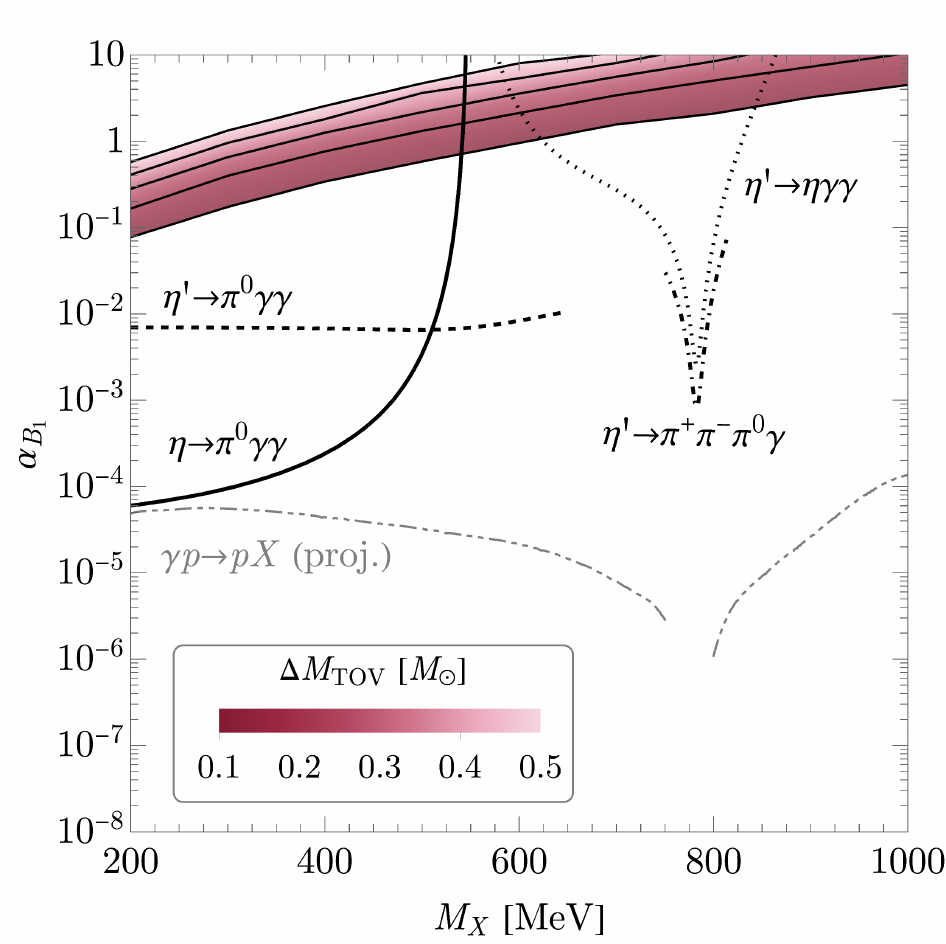}
    \caption{Estimated constraints on the mass and coupling of a $U(1)_{B_1}$ gauge boson from $\eta^{(\prime)}$ decays assuming the SM contribution is zero, albeit disparate nonzero SM assessments exist~\cite{Ametller:1991dp,Bijnens:1995vg,Oset:2002sh,Oset:2008hp,Escribano:2018cwg}. Black lines are based on existing data~\cite{Zyla:2020zbs} after Ref.~\cite{Tulin:2014tya} and the gray line is a projection as noted in text. Also shown is the change to the maximum TOV mass, $\Delta M_\text{TOV} \in [0.1, 0.5]$ $M_{\odot}$; increments of 0.1 $M_{\odot}$ are emphasized in black.
    }
    \label{fig:limit_plot}
\end{figure}

{\it Constraints.}
In Fig.~\ref{fig:limit_plot}, we show the region in the $\alpha_{B_1}$--$M_X$ plane in which $M_\text{TOV}$ is increased by $0.1-0.5 M_\odot$ relative to the APR EoS. We now turn to potential constraints on this scenario from low-energy physics. The presence of NP induces a contribution to the $NN$ scattering lengths. In the Born limit, the $nn$ $^1S_0$ scattering length is modified by
\begin{equation}
    \Delta a_{^1S_0} = \frac{\alpha_{B_1} m_N}{M_X^2} \approx (0.5 \text{ fm}) \times \alpha_{B_1} \left(\frac{\text{600 MeV}}{M_X}\right)^2.
\end{equation}
We emphasize the AV18 potential is phenomenological --- it is fit to low-energy $NN$ data, not derived from first principles. If the potential parameters were determined in the presence of NP, the effects of NP would presumably be obscured; we leave a detailed study to future work~\cite{UsUpcoming}. Therefore, low-energy $NN$ scattering does not provide a robust constraint on this scenario. New baryon-coupled physics can also be probed by lead-neutron scattering \cite{Barbieri:1975xy,Leeb:1992qf}. Ref.~\cite{Leeb:1992qf} presents a constraint for masses below 40 MeV; Ref.~\cite{Tulin:2014tya} extends this into the $\mathcal{O}(100)$ MeV mass range. However, if the range of the new force is not longer than the range of the nuclear force, then it is difficult to disentangle the two without a first-principles description of the latter, and the treatments of Refs.~\cite{Barbieri:1975xy,Leeb:1992qf} are unsuitable to this mass region. As such, we do not consider this constraint further. 

We calculate the contribution of our new boson to several rare $\eta^{(\prime)}$ decays using the vector meson dominance model~\cite{Bando:1984ej,Bando:1985rf,Fujiwara1985nonabelian,Bando:1987br}. We assume that the decays proceed via $\eta^{(\prime)} \to X \gamma$, $X \to \pi^0 \gamma, \, \pi^+ \pi^- \pi^0, \, \eta \gamma$, through $X$-$\omega$ meson mixing and that the SM contribution is zero. The observables are ratios of the rare decay widths to the widths for $\eta^{(\prime)} \to \gamma\gamma$ \cite{Zyla:2020zbs}. The solid curve in Fig.~\ref{fig:limit_plot} shows the constraint derived from $\eta \to \pi^0 \gamma \gamma$. Following Ref.~\cite{Tulin:2014tya}, we require that the contribution from $X$ not exceed $3 \times 10^{-4}$ \cite{Zyla:2020zbs}; equality gives the curve shown. The SM contributions to $\eta^{(\prime)}\to \pi^0 \gamma \gamma$ are not negligible; moreover, different width assessments exist~\cite{Ametller:1991dp,Bijnens:1995vg,Oset:2002sh,Oset:2008hp,Escribano:2018cwg} and further exploration is needed~\cite{UsUpcoming}. The upcoming JLab Eta Factory (JEF) experiment \cite{JEFproposal,JEFproposal2} can perform a bump hunt in $\pi^0\gamma$ invariant mass, greatly enhancing the sensitivity to the $X$ gauge boson while mitigating sensitivity 
to the theoretical production rate~\cite{Gan:2020aco}.

Figure \ref{fig:limit_plot} also shows constraints from decays of $\eta^\prime$ to $\pi^0 \gamma \gamma$ \cite{Ablikim:2016tuo}, $\pi^+ \pi^- \pi^0 \gamma$ \cite{Ablikim:2019wop} and $\eta \gamma \gamma$ \cite{Ablikim:2019wsb}. The possibility of gluonium content in the $\eta^\prime$~\cite{Rosner:1982ey,Kou:1999tt} also complicates their interpretation. Analyses of neutral meson radiative decays do not agree on its size~\cite{Escribano2007JHEP...05..006E,KLOEeta2007PhLB..648..267K}, where the inclusion of $\Gamma(\eta^\prime \to \gamma\gamma)/\Gamma(\pi^0 \to \gamma\gamma)$ data drives this difference and a larger effect~\cite{KLOEeta2007PhLB..648..267K}. Our estimates assume this is zero, so that observed deviations between SM predictions and experiment could also derive from this effect. An alternative strategy for observing $X$ would be to search for bumps in invariant mass distributions in these decays. We caution, however, that there are regions in parameter space in which we expect the $X$ to be wide: this is so for $\alpha_{B_1} \gtrsim \mathcal{O}(10^{-1})$ around the $\omega$ resonance, and for $\alpha_{B_1} \gtrsim \mathcal{O}(1)$ above $M_X \sim 1$ GeV. In these regimes, the $X$ would not present as a localized feature in invariant mass and bump hunts would become less effective. These decays could be measured precisely at JEF and REDTOP \cite{Gatto:2016rae,Gonzalez:2017fku,Gatto:2019dhj}, though the sensitivities have not been benchmarked. Additionally, the ultimate sensitivity of GlueX \cite{Adhikari:2020cvz} to $X$ photoproduction ($\gamma p \to p X$) affords a sensitivity to couplings of order $\mathcal{O}(10^{-5}-10^{-4})$ for narrow $X$ off the $\omega$ resonance \cite{Fanelli:2016utb}; this is also shown in Fig.~\ref{fig:limit_plot}.

{\it Summary.}
We have considered how a new force between first-generation quarks can make NSs for a fixed EoS and many-body method both heavier and puffier. This mechanism has not been considered previously, though new forces for strange quarks have been considered~\cite{Krivoruchenko2009PhRvD..79l5023K}; the attractive $\Lambda$N interaction has long made the existence of $\approx 2M_\odot$ NSs a puzzle~\cite{Demorest:2010bx}, though three-body forces may reduce the effect~\cite{Gerstung:2020ktv}. We have described how our NP scenario can be tested through studies of rare $\eta$ and $\eta'$ decays~\cite{Gan:2020aco, Gatto:2019dhj} and of $X$ photoproduction~\cite{Fanelli:2016utb} at low-energy accelerators. Finally, although we have not resolved the nature of the $\approx 2.6 M_\odot$ compact object in GW190814, this mechanism allows it to more naturally be a NS. The spin of that object, though poorly determined, may have been sufficient to increase its mass by $\sim (0.1-0.4) M_\odot$~\cite{Zhang:2020zsc, Tsokaros:2020hli, Kanakis-Pegios:2020kzp, Cook:1993qr, Lasota:1995eu, Breu:2016ufb, Koliogiannis:2019rvh, Most:2020bba, Essick:2020ghc}; differential rotation can push this even higher~\cite{Baumgarte:1999cq, Gondek-Rosinska:2016tzy, Weih:2017mcw}, but these configurations are not expected to be stable over long time scales. Combining spin effects with NP could yield additional heavy NSs; thus more compact objects in excess of $2 M_\odot$ may eventually be identified, promoting the possibility of new baryonic interactions. 

\begin{acknowledgments}
{\it Acknowledgements.}
We thank the Network for Neutrinos, Nuclear Astrophysics, and Symmetries (\href{https://n3as.berkeley.edu}{N3AS}) for an inspiring environment and support. J.M.B. acknowledges support from the National Science Foundation, Grant PHY-1630782, and the Heising-Simons Foundation, Grant 2017-228, and S.G. acknowledges partial support from the U.S. Department of Energy under contract DE-FG02-96ER40989. 
\end{acknowledgments}

\bibliography{hiddenlives_bib}

\end{document}